\documentclass[paper]{ieice}
\usepackage[pdftex]{graphicx,xcolor}
\usepackage[fleqn]{amsmath}
\usepackage{newtxtext}
\usepackage[varg]{newtxmath}

\field{A}
\title[Spatial Interpolation of RIRs based on DPINNs with RCs]{Spatial Interpolation of Room Impulse Responses based on Deeper Physics-Informed Neural Networks with Residual Connections}
\titlenote{This work was partially supported by JSPS KAKENHI Grant Number 22K12099 and JGC-Saneyoshi Scholarship Foundation. This paper has been submitted to a preprint server~\cite{MYarXiv}.}
\authorlist{%
\authorentry[24fmi09@ms.dendai.ac.jp]{Ken Kurata}{n}{Tokyo Denki University}\MembershipNumber{}
\authorentry[24udc03@ms.dendai.ac.jp]{Gen Sato}{s}{Tokyo Denki University}\MembershipNumber{2407779}
\authorentry[itsunokuni@mail.dendai.ac.jp]{Izumi Tsunokuni}{m}{Tokyo Denki University}\MembershipNumber{2326481}
\authorentry[yusuke.ikeda@mail.dendai.ac.jp]{Yusuke Ikeda}{m}{Tokyo Denki University}\MembershipNumber{2407766}
}
\affiliate[Tokyo Denki University]{Tokyo Denki University, Tokyo, 120--8551 Japan}

\begin{document}
\maketitle
\begin{summary}
The room impulse response (RIR) characterizes sound propagation in a room from a loudspeaker to a microphone under the linear time-invariant assumption.
Estimating RIRs from a limited number of measurement points is crucial for sound propagation analysis and visualization. Physics-informed neural networks (PINNs) have recently been introduced for accurate RIR estimation by embedding governing physical laws into deep learning models; however, the role of network depth has not been systematically investigated.
In this study, we developed a deeper PINN architecture with residual connections and analyzed how network depth affects estimation performance.
We further compared activation functions, including tanh and sinusoidal activations. Our results indicate that the residual PINN with sinusoidal activations achieves the highest accuracy for both interpolation and extrapolation of RIRs. Moreover, the proposed architecture enables stable training as the depth increases and yields notable improvements in estimating reflection components. These results provide practical guidelines for designing deep and stable PINNs for acoustic-inverse problems.
\end{summary}
\begin{keywords}
Room acoustics, room impulse response estimation, sound field reconstruction, SIREN, residual connections
\end{keywords}

\section{Introduction}
\label{sec:introduction}
The room impulse response (RIR) is a signal that represents the sound propagation in a room from a loudspeaker to a microphone, based on the assumption that the sound field is linear and time invariant. The RIR is utilized for various acoustic fields and applications, including sound field reproduction~\cite{sound field reproductions01,sound field reproduction02} and sound source localization~\cite{sound field Localization01,sound field Localization02}. Moreover, the early part of the RIR significantly influences the sound localization and timbre impression of the sound source~\cite{sound field teii}.  

In applications where spatial variations in sound propagation are essential---such as sound field visualization~\cite{sound field visualization01,sound field visualization02,visUchida} and sound field control~\cite{ANC01,ANC02,SFSuzuki}---acquiring RIRs over large regions at high spatial densities is crucial for accurately representing the sound field.
In particular, the early part of RIR can vary substantially with the measurement position.
However, when RIR measurements are required over large regions or at high spatial densities, it is impractical---in terms of the measurement cost---to physically place microphones at all desired locations. Therefore, various methods have been proposed to estimate the RIRs in a region of interest using measurements from only a limited number of microphones.

One well-known category of approaches is  model-based approaches that incorporate compressed sensing (CS)~\cite{CS01,CS02}. In CS-based methods, the sound field is represented as a sparse linear combination of physical models, such as plane waves~\cite{planewave01} and point sources~\cite{ESM01}.

For example, the sparse equivalent source method~\cite{CESM01,CESM02} represents the sound field by superposition of sparse point sources. Additionally, in the study by Verburg {\it et al.}, the sound field was modeled as a superposition of plane waves, assuming that the weights of this superposition were sparse~\cite{planewave01}. Furthermore, Koyama {\it et al.} reconstructed the sound field by modeling direct sound with sparse point sources and the reverberant component using a combination of sparse plane waves and a low-rank matrix~\cite{planewaveCESM01}.

Another line of research focuses on least-squares-based solutions utilizing kernel ridge regression (KRR)~\cite{KRR01,KRR02,KRR03}. Horiuchi {\it et al.} proposed a method for automatically learning and optimizing the parameters of the kernel function used in sound field estimation directly from  measured signals~\cite{KRR01}. Ueno {\it et al.} reported that estimation accuracy was improved by incorporating the Helmholtz equation into the KRR framework~\cite{KRR02}. Additionally, in the study by Ribeiro {\it et al.}, 
the kernel is defined as the sum of two components: one kernel is designed to capture direct sound and early reflections, and the other to capture reverberation~\cite{KRR03}.

RIR estimation methods based on deep learning have been rapidly developed in recent years. Convolutional neural networks (CNNs)~\cite{CNN01} and generative adversarial networks (GANs)~\cite{GAN01} can model the relationship between inputs and outputs, and have been applied to estimate RIRs~\cite{CNN02,GAN02}. Lluís {\it et al.} proposed a U-Net architecture~\cite{U-Net} trained on the sound field data obtained using the Green's function in a rectangular room as a super-resolution task~\cite{CNN02}. Fernandez-Grande {\it et al.} employed a GAN-based approach to estimate RIRs from limited measurement signals, achieving improved estimation accuracy compared to previous plane wave regression methods~\cite{GAN02}.

However, these data-driven deep learning models require a substantial amount of training data. To address this challenge, physics-informed neural networks (PINNs)~\cite{PINN01} have been introduced that incorporate physical laws into their loss functions.
A key advantage of PINNs is that they learn an implicit neural representation of a specific sound field by incorporating physical laws, including boundary and initial conditions, rather than learning data structure from large datasets.
Furthermore, by constraining network outputs to adhere to the governing partial differential equations (PDEs) that represent physical laws, PINNs are expected to achieve a higher estimation accuracy in solving forward and inverse problems than other deep learning methods.
For example, in forward problems, Raissi {\it et al.} applied PINNs to the Schrödinger and Navier-Stokes equations~\cite{PINN01}. Moreover, Alkhadhr {\it et al.} reported that for the linear wave equation, PINNs achieved comparable accuracy to the finite difference method with approximately half the computation time~\cite{PINN02}.

Furthermore, several studies have applied PINNs to inverse problems, including estimating RIRs at arbitrary positions from measured RIRs~\cite{PISIREN01,PISIREN02,PISIREN03,PISIREN04}.
In particular, sinusoidal representation networks (SIREN)~\cite{SIREN01}, which employ sinusoidal activation functions in a multilayer perceptron (MLP)~\cite{MLP}, are one of the commonly used architectures in PINN-based RIR estimation.
SIREN is an effective architecture for learning neural implicit representations of various signals.
Specifically, Pezzoli {\it et al.}  compared the PINNs and CS methods and showed that the PINNs method achieved an improvement in RIR estimation accuracy~\cite{PISIREN02}. Furthermore, Olivieri {\it et al.} demonstrated a three-dimensional sound field reconstruction using PINNs~\cite{PISIREN04}. In these studies, the experiments were conducted using relatively shallow architectures with a limited number of hidden layers.

Increasing the number of network layers generally improves the representational capacity, albeit at the cost of longer training time. This increased capacity can be beneficial in challenging scenarios such as long signals or noisy environments, where shallow networks may not achieve sufficient estimation accuracy.
In PINNs, deeper networks often suffer from training instabilities including exploding or vanishing gradients~\cite{PINNsRes01}. Consequently, simply increasing the number of layers does not necessarily improve the performance and may even degrade the estimation accuracy.

In recent years, various architectural improvements have been proposed to facilitate deeper neural networks, including residual and densely connected architectures~\cite{ResNet01,DenseNet01}.
In the context of PINNs, these ideas have been introduced mainly for forward problems to stabilize the training and enable deeper networks~\cite{PINNsRes02,PINNsRes04,PINNsRes03}.
Cheng and Zhang introduced ResNet-like blocks into PINNs for fluid dynamics forward problems~\cite{PINNsRes02}. ResNet-style skip connections improve trainability by enabling residual learning through additive identity mappings~\cite{ResNet01}. 
Jiang \textit{et al.} demonstrated that residual connections combined with element-wise multiplication facilitate deeper PINNs and improve the accuracy of forward problems, such as one-dimensional wave propagation~\cite{PINNsRes04}.

The application of residual connections to PINNs has also been explored to solve inverse problems. For instance, Tian {\it et al.} performed wind-field reconstruction using PINNs with an MLP augmented by residual connections~\cite{PINNsRes01}. This method achieved robust estimation under noisy conditions, and by mitigating the vanishing gradient problem, it enabled the accurate computation of the governing physical laws. However, it should be noted that the model employed in their study was not a SIREN-based MLP, but rather a standard MLP utilizing a tanh activation function.

In addition, residual connections in SIREN-based architectures have been studied to facilitate deeper networks. Lu {\it et al.} demonstrated that adding residual connections to SIREN without physical laws stabilized the training and improved the accuracy of the 3D image reconstruction~\cite{SIRENRes01}. Based on this architecture, we previously proposed relatively shallow PINNs incorporating SIREN with residual connections for 2D sound fields~\cite{PISIRENRes01}. This previous model employed a compact architecture consisting of only 4 hidden layers with 256 neurons per layer. Consequently, the impact of hidden layer depth on estimation accuracy was not verified in that study, nor was estimation performed in a three-dimensional sound field.

Therefore, for inverse problems in PINN-based RIR estimation, the effect of network depth in architectures with residual connections has not been systematically studied. In addition, the relative suitability of activation functions (e.g., tanh vs. sin) for this acoustic-inverse problem remains unclear because direct comparisons are limited. Consequently, it is not yet well understood how increasing depth influences RIR estimation performance and which activation function is most appropriate for the task.

In this study, we investigate the acoustic-inverse problem of RIR estimation in a three-dimensional sound field to address these open questions. To this end, we consturuct a deeper PINN architecture that combines SIREN with residual connections. 
Furthermore, we quantitatively compare the effects of activation functions, network depth, and residual connections, providing insight into the design of deep and stable PINN for acoustic inverse problems.

The main contributions of this paper are summarized as follows:
\begin{itemize}
  \setlength{\itemsep}{0.2em}
  \setlength{\parskip}{0pt}
  \setlength{\parsep}{0pt}
  \item \textbf{Systematic depth study for 3D RIR PINNs:} We conduct a controlled investigation of network depth in PINNs for 3D RIR estimation, clarifying when deeper models degrade or improve performance under physics-based training.
  \item \textbf{Residual SIREN tailored for physics-informed training:} We propose a deep physics-informed architecture that combines SIREN activations with residual connections (PINNs-SIREN-Res) to stabilize training in deep settings and to improve reconstruction accuracy for RIR estimation.
  \item \textbf{Task-oriented evaluation and insights:} Through extensive comparisons (tanh vs.\ sin, with/without residuals), we provide practical design insights validated on interpolation and extrapolation, temporal behavior, and frequency-band performance.
\end{itemize}

The remainder of this paper is organized as follows. Section II explains the problem formulation. Section III introduces the theoretical background and the architecture of the proposed method. Section IV outlines the experimental conditions used to validate the proposed method. Section V demonstrates the effectiveness of the proposed method through simulation results, focusing on the estimation accuracy in both time and frequency domains. Finally, Section VI concludes the paper.

\begin{figure}[tb]
\centering
\includegraphics[width=0.9\linewidth]{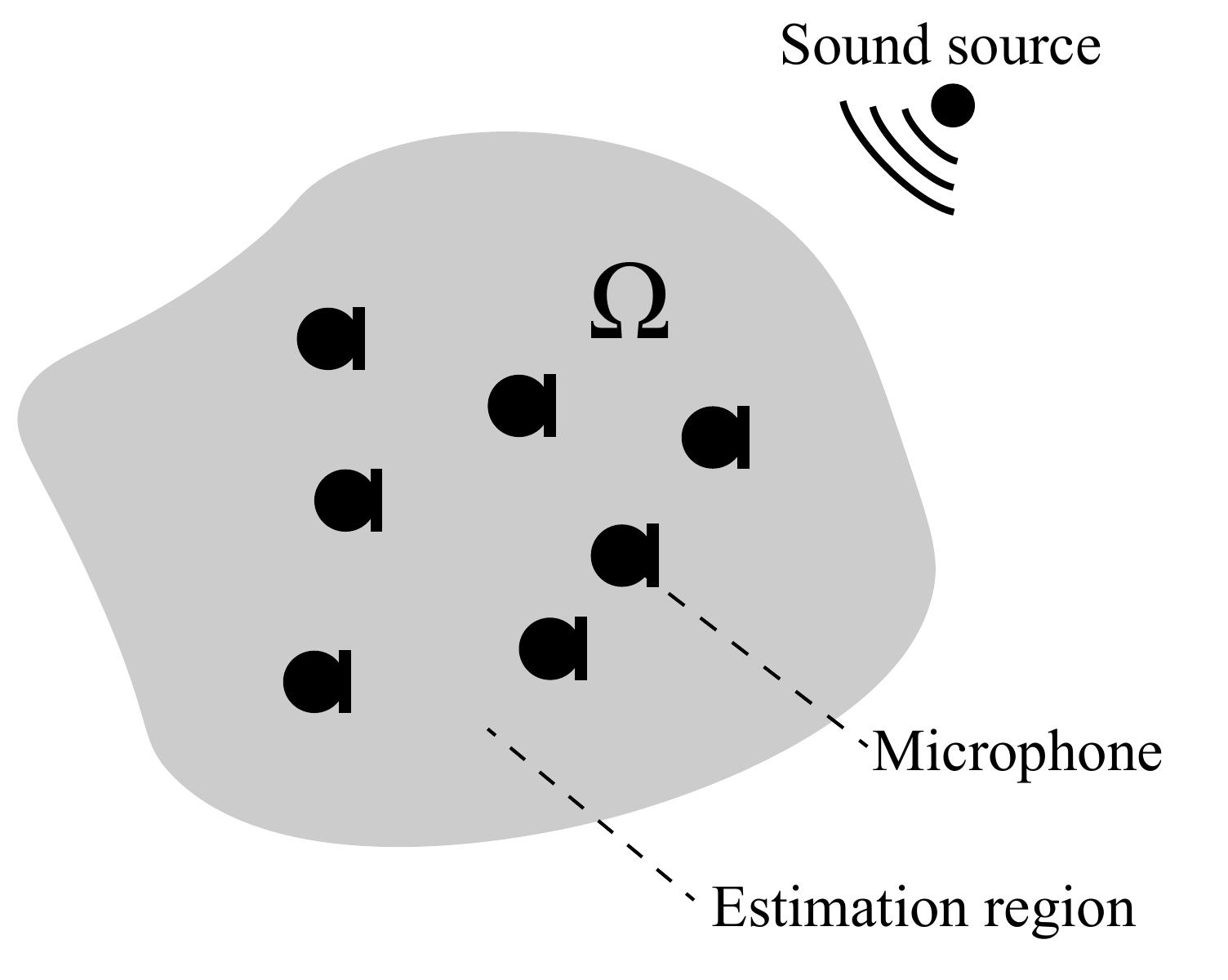}
\vspace{10pt}
\caption{Problem of RIR estimation from a limited number of microphones}
\vspace*{-3pt}
\label{SF_prob}
\end{figure}

\begin{figure*}[tb]
\centering
\includegraphics[width=\linewidth]{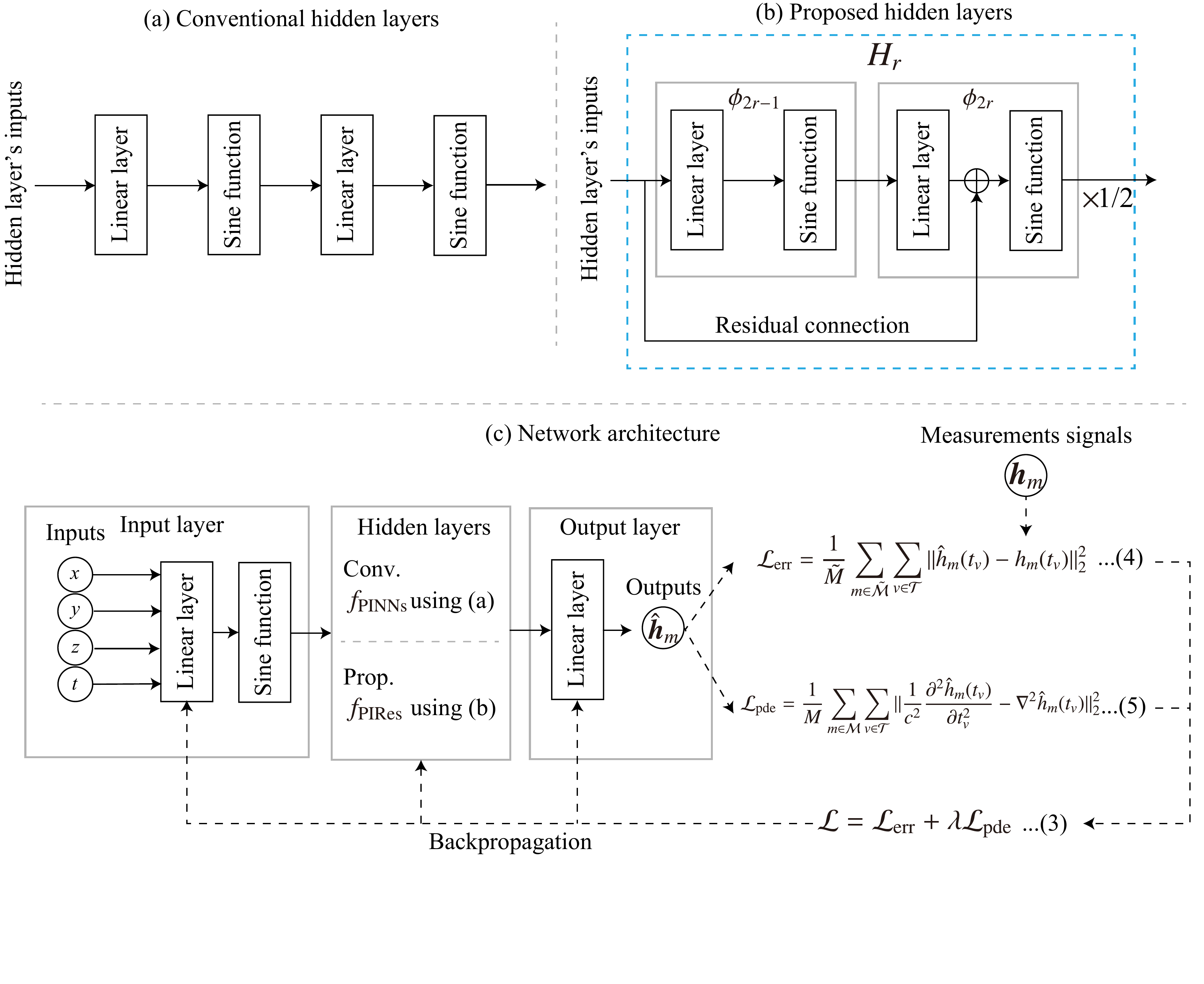}
\vspace{10pt}
\caption{(a) Conventional hidden layers architecture. (b) Proposed  hidden layers architecture. (c) Network architecture. Inputs are coordinates $(x,y,z)$ and time $t$. The network is composed of $N$ hidden layers. The loss function is calculated by the output RIRs ${{\boldsymbol{\hat{h}}_m}}(m\in\mathcal{M})$ and measured RIRs ${\boldsymbol{h}_m}$. The network parameters ($\mathbf{w}$ and $\mathbf{b}$) are then optimized by minimizing this total loss function, Eq.~(\ref{PI-SIREN_loss}), via backpropagation. $\mathcal{M}$ is the set of indices for the estimation points.}
\vspace*{-3pt}
\label{Networks}
\end{figure*}

\section{Problem Formulation}

In this study, we estimate the RIRs within a three-dimensional region $\Omega(\subset \mathbb{R}^3)$ that contains neither sound sources nor diffracting objects, as shown in Fig.~\ref{SF_prob}.
Under these assumptions, the RIR $h({\mathbf x},t)$ at arbitrary position ${\mathbf x}(\in \Omega$) satisfies the following homogeneous wave equation:
\begin{equation}
    \frac{1}{c^2}\frac{\partial^2 h({\mathbf{x}}, t)}{\partial t^2} - \nabla^2 h({\mathbf{x}}, t) = 0, 
    \label{eq:wave}
\end{equation}
where $c$ is the speed of sound and $\nabla$ is the gradient.

Let $h({\mathbf x}_m,t)$ denote the RIR measured at $m$-th microphone position ${\mathbf x}_m (\in \Omega)$ in time $t$. 
We estimate the RIR signal by fitting the measured signals $h({\bf x}_m,t)$ at the measurement points ${\bf x}_m$ while enforcing the wave equation at arbitrary points in $\Omega$. Accordingly, the RIR at an arbitrary position ${\bf x}$ can be expressed as
\begin{equation}
    \hat{h}({\mathbf x}, t) = \mathcal{U}({\mathbf x}, t)
\end{equation}
where the function $\mathcal{U}(\cdot)$ denotes a neural network trained using both the measurement loss and the physics-based constraint in Eq.~(\ref{eq:wave}).

Although a room impulse response is generated by an impulsive source and boundary reflections, our reconstruction is formulated on a spatial region $\Omega$ that does not include the source position or any obstacles. Hence, within $\Omega$, the field after excitation satisfies the source-free wave equation, while the influence of the source and room boundaries is implicitly embedded in the observed microphone signals.

In the following sections, both the measurement and evaluation positions are discretized. Accordingly, the RIR at the $m$-th position is denoted by $h_m(t)$ rather than $h({\mathbf x}_m,t)$, where the subscript $m$ indexes the spatial location.

\section{method}
This section provides an overview of PINNs and the proposed method for RIR estimation, specifically to estimate unobserved RIRs from a limited set of measured RIRs. 
PINNs are well suited to sound-field interpolation because they embed governing equations in the loss function, enabling accurate estimation in the neighborhood of measurement points from limited data.

\subsection{Physics-informed neural networks (Conventional Method)}
PINNs represent a deep learning methodology that incorporates physical laws into the loss function~\cite{PINN01}. Unlike conventional data-driven neural networks, PINNs leverage physical laws to guide the learning process and ensure physically consistent predictions. 
In acoustics, the wave equation Eq.(~\ref{eq:wave}) governs the sound propagation. We incorporate this into the loss function and compute the required second-order derivatives via automatic differentiation.

In the following sections, we describe the conventional PINN framework, first presenting the loss function, and then the network architecture.

\subsubsection{Loss Functions}

For the acoustic-inverse problem, the PINN is trained by minimizing the following loss function, which consists of a data-misfit term at the microphone locations and a governing-equation (wave-equation) residual term:
\begin{equation}
    \label{PI-SIREN_loss}
    {\mathcal L} = {{\mathcal L}_{\mathrm{err}}} + \lambda {{\mathcal L}_{\mathrm{pde}}},
\end{equation}
\begin{equation}
   \label{PI-SIREN_loss_err}
     {\mathcal L}_{\mathrm{err}} =
     \frac{1}{\tilde{M}}\sum_{m\in \tilde{\mathcal M}}\sum_{v\in {\mathcal T}}\|{{{\hat{h}}_m}(t_v)}-{{h}_m(t_v)}\|_2^2,
\end{equation}
\begin{equation}
    \label{PI-SIREN_loss_pde}
      {\mathcal L}_{\mathrm{pde}} = \frac{1}{M}\sum_{m\in {\mathcal M}}\sum_{v\in {\mathcal T}}\| \frac{1}{c^2}{\frac{\partial^2{{{\hat{h}}}_m(t_v)}}{\partial t_v^2}}-\nabla^2{{\hat{h}}_m(t_v)}\|_2^2.
\end{equation}    
Here, $\mathcal{L}_{\mathrm{err}}$ is defined as the mean squared error (MSE), which quantifies the estimation error between measured signals $h_m$ and estimated signals ${\hat{h}_m}$. 
$\mathcal{T}$ denotes the set of time indices, where $t_v$ represents the $v$-th time sample. 
$\mathcal{L}_{\mathrm{pde}}$ represents the wave equation residual term.
As ${\mathcal L}_{\mathrm{pde}}$ approaches zero, the estimated signals ${\hat{h}_m}$ increasingly satisfy the physical constraints imposed by the wave equation.
The balancing parameter $\lambda$ adjusts the trade-off between the data fidelity and physical consistency. 
$\tilde{\mathcal M}$ is the set of position indices for the measurement positions and $\mathcal{M}$ is the set of position indices for the estimation points. $\tilde{M}=|\tilde{\mathcal M}|$ and $M=|{\mathcal M}|$ indicate the number of measurement positions and estimation points, respectively.

\subsubsection{Network Architecture}
Standard PINNs employ an MLP as an implicit function representation, mapping the spatial coordinates to target field variables~\cite{PINN01}.
This coordinate-based representation provides a mesh-free, continuous approximation of the target field and enables efficient evaluation of the PDE residual via automatic differentiation.

In this study, the wave-equation residual is computed from the second-order derivatives of the network output.
However, common activation functions, such as ReLU and tanh, can be unsuitable when accurate high-order derivatives are required~\cite{SIREN01}. In particular, ReLU yields piecewise-linear outputs whose second and higher-order derivatives are zero almost everywhere. Although tanh is smooth, its derivatives can rapidly diminish and may attenuate fine-scale variations during repeated differentiation. 
To address this issue, we utilize the SIREN~\cite{SIREN01}, an MLP-based architecture that uses a sine function as an activation function, enabling more accurate computation of higher-order derivatives. 

The hidden layers of the SIREN network $f_\mathrm{PINNs}$ can be expressed as a composite function of each network layer $\phi_n$ as follows:
\begin{equation}
\label{eq:siren}
 f_\mathrm{{PINNs}}({\mathbf w},{\mathbf x}) = (\phi_N \circ \phi_{N-1}\circ\cdots \circ\phi_1)({\bf x})
\end{equation}
\begin{equation}\label{eq:sirenphi}
 \phi_n({\mathbf x}_n) = \sin{(\omega_0({\mathbf x}_n{\mathbf w}_n+{\mathbf b}_n))}
\end{equation}
where $\omega_0$, $\mathbf{w}_n$, and $\mathbf{b}_n$ are the hyperparameters, network weights for the $n$-th layer, and biases, respectively, and ${\mathbf x}_n$ is the input to the $n$-th layer, consisting of spatial coordinates and time. $N$ denotes the total number of hidden layers $\phi$. 
Fig.~\ref{Networks}(a) shows the conventional model architecture. The conventional PINNs architecture consists of a stack of fully-connected (linear) layers and activation functions.

Following Sitzmann {\it et al.}~\cite{SIREN01}, to stabilize the training in the SIREN architecture, the input is normalized to the range $[-1, 1]$, and the weights of each layer are initialized as follows:
\begin{equation}\label{eq:w1}
\mathbf{w}_0 \sim \mathcal{U}(-1/s_0, 1/s_0),
\end{equation}
\begin{equation}\label{eq:w2}
\mathbf{w}_{n} \sim \mathcal{U}(-\frac{\sqrt{6/s_{n}}}{\omega_0},\frac{\sqrt{6/s_{n}}}{\omega_0}),\quad n=1,2,3,...,N
\end{equation}
where $s_0$ is the input dimension, and $s_{n}$ is the number of neurons in the $n$-th hidden layer.

\subsection{PINNs with residual connections (Proposed Method)}
In general, MLP architectures, including SIREN, are expected to enhance their expressive power as network depth increases. However, simply increasing the number of layers often results in unstable training owing to vanishing and exploding gradients.

To address this issue and achieve stable network training, this study incorporates residual connections into the SIREN. Specifically, we adopt a block structure similar to ResNet-18~\cite{ResNet01}, which enables deepening by preserving the input information during the learning process. However, to comply with the SIREN requirement that inputs to hidden layers remain within $[-1, 1]$, we introduce additional scaling. After adding the input and output of the hidden layer, we multiply the result by $1/2$ to maintain the values within this range.

As shown in Fig.~\ref{Networks}(b), the hidden layers $f_\mathrm{PIRes}$ can be expressed as a composite function of the network layers $\phi_n$, as defined in Eq.~(\ref{eq:sirenphi}) as follows:
\begin{equation}
    \label{SIRENRes_arc}
    f_\mathrm{{PIRes}}({\mathbf{w,x}}) = (H_{{R}} \circ H_{R-1}\circ \cdots  H_{{r}} \circ \cdots H_1)(\mathbf x),\\
\end{equation}
\begin{equation}
    \label{Liner}
    H_{r}\!\!=\!\!\frac{1}{2}\sin( \omega_0(\phi_{2r-1}({\mathbf{x}}_{2r-1})\mathbf{w}_{2r} \!+\! \mathbf{b}_{2r})\!+\!{\mathbf{x}}_{2r-1}) 
\end{equation}
where $H_{r}$ is the $r$-th residual block $(r=1,\dots,R)$, where $R=N/2$. Parameters $\mathbf{w}$ and $\mathbf{b}$ are the weights and biases of the network, respectively. 
$\omega_0$ is a hyperparameter for sine activation, as defined in Eq.~(\ref{eq:sirenphi}).

As part of the network and data initialization procedure, to ensure the training stability of this network, the input data are normalized to the range $[-1, 1]$, and the weights in each layer are initialized according to Eqs.~(\ref{eq:w1}) and (\ref{eq:w2}), respectively.

Although SIRENs with residual connections have been proposed in conventional studies~\cite{SIRENRes01}, our architecture is distinguished by a different input-injection point, namely, we add the input at a different stage of the network.

\subsection{Training Processes}
A model is constructed to estimate RIRs using the networks presented in Eqs.~(\ref{eq:siren}), (\ref{eq:sirenphi}), (\ref{SIRENRes_arc}) and (\ref{Liner}). The network architectures of the conventional and proposed methods are illustrated in Fig.~\ref{Networks}(c). It should be noted that although the conventional and proposed methods share the same training procedure, they differ in their underlying network architectures.

First, during the training phase, the inputs to the model consist of discretized time steps combined with two types of spatial coordinates. The first type is the coordinates of the measurement points, which are required to calculate Eq.~(\ref{PI-SIREN_loss_err}). The second type comprises the coordinates of $\tilde{M}$ randomly selected points within the estimation region, which are required to calculate Eq.~(\ref{PI-SIREN_loss_pde}). These inputs can be represented as follows:
\begin{equation}
    \label{eq:inputs}
    {\mathbf{x}_\mathrm{meas}} = [t_v,x_{\tilde{m}},y_{\tilde{m}},z_{\tilde{m}}] \quad\tilde{m}\in{\mathcal{\tilde{M}}}, {v}\in{\mathcal{T}},
\end{equation}
\begin{equation}
    \label{eq:inputs_rand}
    {\mathbf{x}^{(k)}_\mathrm{est}} = [t_v,x^{(k)}_m,y^{(k)}_m,z^{(k)}_m]\quad{m}\in{\mathcal{{M}}}, {v}\in{\mathcal{T}},
\end{equation}
where ${\mathbf{x}_\mathrm{meas}}$ is the input used to compute Eq.~(\ref{PI-SIREN_loss_err}), which consists of the discretized time sample $t_v$ and the coordinates of the measurement point ($x_{\tilde{m}}$, $y_{\tilde{m}}$, $z_{\tilde{m}}$). 
The other input, ${\mathbf{x}^{(k)}_\mathrm{est}}$, is used to compute Eq.~(\ref{PI-SIREN_loss_pde}), and consists of the discretized time sample $t_v$ and the coordinates of the estimation point ($x_{{m}}$, $y_{{m}}$, $z_{{m}}$).
The estimation coordinates are updated at each iteration, where $k$ is the iteration index.

Both inputs, ${\mathbf{x}_\mathrm{meas}}$ and ${\mathbf{x}^{(k)}_\mathrm{est}}$, are fed into the network. The outputs $\hat{h}_m$ denote the RIRs at the discretized time sample and spatial coordinates. The MSE, Eq.~(\ref{PI-SIREN_loss_err}), is calculated using the output $\hat{h}_m$ corresponding to $\mathbf{{x}_{\mathrm{meas}}}$. In addition, the wave equation is computed using the output $\hat{h}_m$ corresponding to ${\mathbf{x}^{(k)}_\mathrm{est}}$. To compute the wave equation, the temporal and spatial derivatives of output $\hat{h}_m$ corresponding to ${\mathbf{x}^{(k)}_\mathrm{est}}$ are calculated.

Backpropagation is then performed using the total loss function $ {\mathcal L}$, Eq.~(\ref{PI-SIREN_loss}), which balances the MSE and wave equation loss. Thus, by incorporating the wave equation into the loss function alongside the MSE, the PINNs constrain the network output to adhere to the governing physical law. 
By enforcing the governing physical law through this wave equation loss ${\mathcal L}_{\mathrm{pde}}$, the model can estimate the RIRs in the vicinity of the measurement points and suppress noise in the measured RIRs.

\begin{figure}[tb]
\centering
\includegraphics[width=0.9\linewidth]{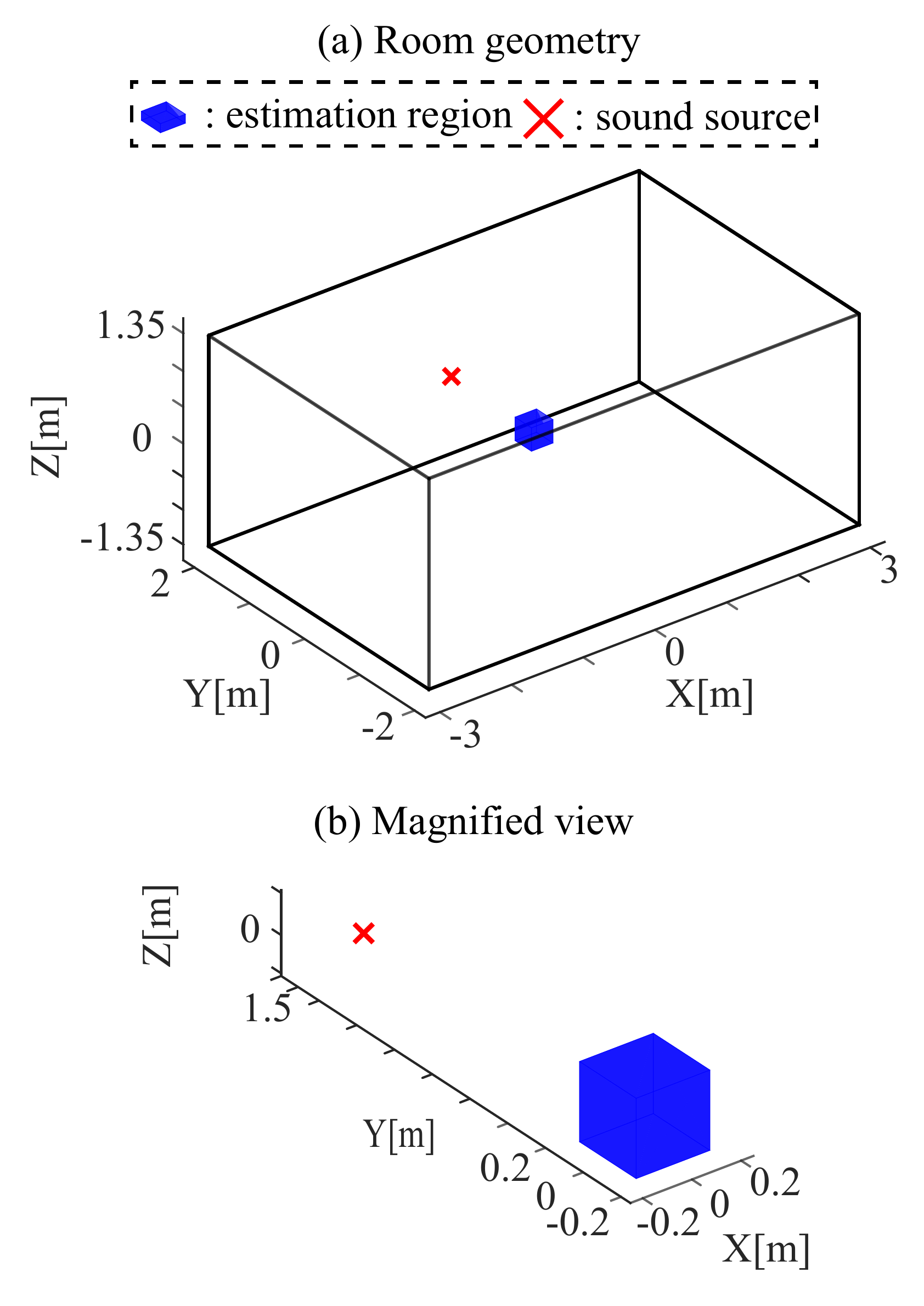}
\vspace{10pt}
\caption{(a) Room geometry. (b) Magnified view of the estimation region and source position within the room. The sound source is positioned $1.5$ m from the room's center in the positive y-direction. The estimation region is a $0.3$ m cube, centered in the room. Microphones are placed within this estimation region, and the RIRs for this entire region are estimated from the measured signals. The evaluation points are defined by discretizing the estimation region into a $14 \times 14 \times 14$ grid.}
\vspace*{-3pt}
\label{rooms}
\end{figure}

\begin{figure}[tb]
\centering
\includegraphics[width=0.8\linewidth]{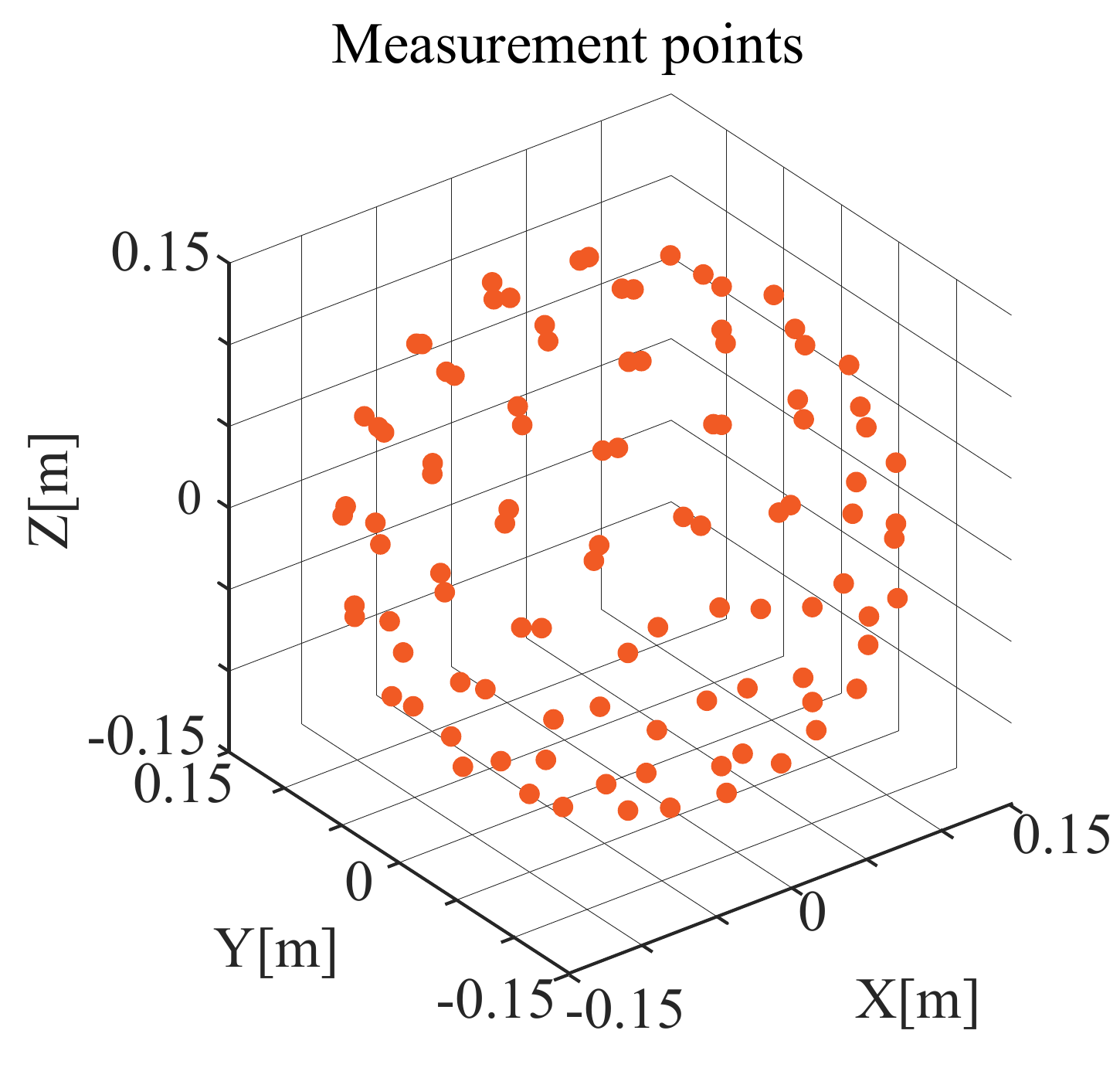}
\vspace{10pt}
\caption{Arrangement of the measurement points. The points are positioned equidistantly with a $0.05$ m spacing on a sphere with a radius of $0.15$ m. The center of the sphere coincides with the center of the room.}
\vspace*{-3pt}
\label{micarray}
\end{figure}

\section{Experimental Setup}
Simulation experiments were conducted to estimate the early part of RIRs from a limited number of microphone signals in a three-dimensional sound field. The experiments aimed to examine the impact of residual connections and network depth on estimation accuracy and to select an appropriate activation function.
The estimation accuracies of the four methods were compared: three conventional methods---PINNs with SIREN (PINNs-SIREN), PINNs with the hyperbolic tangent function (PINNs-tanh), PINNs with the hyperbolic tangent function and residual connections (PINNs-tanh-Res)---, and PINNs incorporating SIREN with residual connections (PINNs-SIREN-Res, the proposed method). All the architectures were implemented using PyTorch (version 1.12.0).

\begin{figure*}[tb]
\centering
\includegraphics[width=1\linewidth]{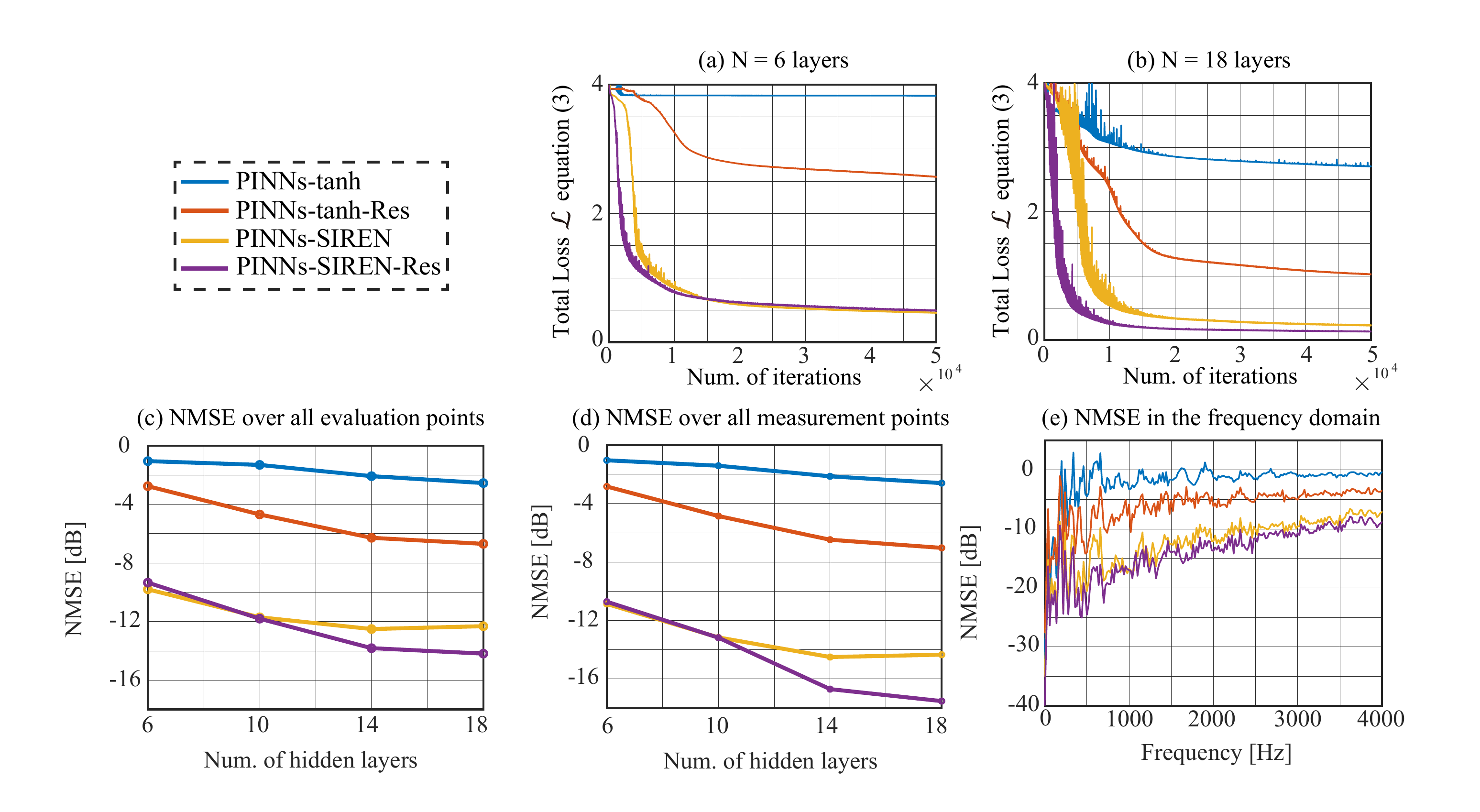}
\vspace{10pt}
\caption{The trajectory of the loss function and estimation accuracy of evaluation and measured points. The trajectory of the loss function $\mathcal{L}$ over iterations for models with (a) $N=6$ and (b) $N=18$ hidden layers. (c) NMSE of evaluation points versus the number of hidden layers. (d) NMSE of measurement points versus the number of hidden layers. (e) Frequency-domain NMSE for the best-performing model of each method, where the network depth is selected to maximize performance.}
\vspace*{-3pt}
\label{totalloss}
\end{figure*}

\subsection{Simulation Setup}
Figs.~\ref{rooms} and \ref{micarray} show the room geometry, estimation region, and arrangement of the measurement points used in this experiment.

As shown in Fig.~\ref{rooms}, the room geometry for the simulation was $6.0 \times 4.0 \times 2.7 \ {\mathrm m}^3$, and a point source was placed at (0, 1.5, 1.35). Following the conditions reported in a publicly available RIR dataset~\cite{MeshRIR}, the wall reflection coefficient was set to achieve a $T_{60} = 0.38$ s. The estimation region was $0.3\times0.3\times0.3 \ {\mathrm m}^3$ and was positioned at the center of the room.

The simulated signals had a length of 400 samples and a sampling frequency of 8000 Hz. To simulate realistic measurement conditions, white noise was added to the measured signals to achieve a signal-to-noise ratio (SNR) of 20 dB.
RIRs were simulated using the image source method~\cite{pyroomaco} with the SFS-toolbox~\cite{SFStoolbox}, considering reflections up to the seventh order based on these room conditions and signal length.

As shown in Fig.~\ref{micarray}, the measurement points were arranged equidistantly on a spherical surface. The total number of microphones was $M = 100$, with a radius of $0.15$ m, and the microphone spacing was approximately $0.05$ m. The sphere was centered at the room center, which was considered as the origin $(0, 0, 0)$.

\subsection{Network parameters}
For all methods, we constructed models with $N = 6, 10, 14$ and $18$ layers using  $s_{n} = 256$ neurons in each hidden layer. By comparing networks of different depths, we evaluated the effect of depth and assessed whether deeper architectures improve performance relative to shallower ones.

These networks were trained for $50,000$ iterations using the Adam optimizer. At each iteration, $\tilde{M} = 100$ random points were sampled for training. The learning rate was initialized at $1.0 \times 10^{-4}$ and exponentially decayed by a factor of 0.98 every 100 iterations, to a final value of $1.0 \times 10^{-6}$.

The hyperparameter $\omega_0$ and balance parameter $\lambda$ were set to $\omega_0=7$ and $\lambda = 1.0\times 10^{-6}$, respectively. Note that $\omega_0$ and weight initialization (defined in Eqs.~(\ref{eq:w1}) and (\ref{eq:w2})) were only applied to the PINNs-SIREN and PINNs-SIREN-Res models.

The training of all methods was performed on a workstation equipped with an AMD EPYC 7543P CPU (32 cores, 64 threads, 2.8 GHz), an NVIDIA RTX A6000 GPU (48 GB VRAM), and 256 GB of RAM.

\subsection{evaluation method}
The estimation region was discretized into a $14 \times 14 \times 14$ grid, yielding $M_{\rm eval}= 2744$ evaluation points. The coordinates of grid points were fed into each trained model to estimate the RIR at the corresponding positions. The estimation accuracy was evaluated using the normalized mean squared error (NMSE), defined as follows:
\begin{equation}\label{eq:NMSE}
 {\mathrm{NMSE}} = 10\log_{10}\sum_{m = 1}^{M_{\rm eval}}\sum_{v = 1}^{T}\frac{\|h_m(t_v) - \hat{h}_m(t_v)\|^2_2}{\|h_m(t_v)\|^2_2}.
\end{equation}
$h_m(t_v)$ and $\hat{h}_m(t_v)$ denote $m$-th ground truth and $m$-th estimated RIR, respectively. $v(=1,\dots,T=400)$ denotes time index. The evaluation was conducted using noise-free RIRs as the ground truth to compute the NMSE.

\section{Results}

\begin{table}[tb]
\centering
\caption{NMSE [dB] of the lowest-NMSE model across the tested depths for each method with $50,000$ training iterations. This table presents the Normalized Mean Squared Error (NMSE) values achieved after $50,000$ iterations for each model (PINNs-tanh, PINNs-tanh-Res, PINNs-SIREN, and PINNs-SIREN-Res).}
\setlength{\tabcolsep}{3pt}
\begin{tabular}{c|c|c|c}
\hline
PINNs-tanh& 
PINNs-tanh-Res& 
PINNs-SIREN&
PINNs-SIREN-Res\\
\hline
$-2.6$&
$-6.7$&
$-12.5$&
$-14.2$\\
\hline
\end{tabular}
\label{NMSE}
\end{table}
\subsection{Evaluation of overall estimation accuracy}

In this section, we evaluate the proposed method in terms of training stability and estimation accuracy over the entire estimation domain.

\subsubsection{Training Stability and Convergence}
First, we examine the training stability and the convergence of models. Figs.~\ref{totalloss}(a) and (b) illustrate the trajectories of the total loss function $\mathcal{L}$ over training iterations for models with $N=6$ and $N=18$ hidden layers, respectively. 

As shown in Fig.~\ref{totalloss}(a), the total loss function converges successfully for the $N=6$ model. 
Regarding the activation functions, PINNs-SIREN and PINNs-SIREN-Res converged more rapidly and stably than PINNs-tanh and PINNs-tanh-Res. Overall, the SIREN-based models showed consistently better convergence behavior than the tanh-based models, suggesting that SIREN can be advantageous for modeling acoustic wave propagation in this setting.

\begin{table*}[tb]
\centering
\caption{NMSE [dB] in each frequency band for the evaluated models.}
\setlength{\tabcolsep}{3pt}
\begin{tabular}{c|c|c|c|c}
\hline
Frequency Range [Hz]&
PINNs-tanh& 
PINNs-tanh-Res& 
PINNs-SIREN&
PINNs-SIREN-Res\\
\hline
$0-1000$&
$-7.2$&
$-13.2$&
$-21.2$&
$-23.7$\\
$1000-2000$&
$-1.3$&
$-5.5$&
$-12.7$&
$-14.6$\\
$2000-3000$&
$-1.0$&
$-4.8$&
$-10.2$&
$-12.0$\\
$3000-4000$&
$-0.8$&
$-3.9$&
$-8.1$&
$-9.4$\\
\hline
\end{tabular}
\label{NMSE_fq_block}
\end{table*}

Within each activation function, incorporating residual connections led to faster convergence. This suggests that residual connections can accelerate training and improve convergence behavior even for shallow networks.
In particular, comparing PINNs-SIREN with PINNs-SIREN-Res, the residual variant achieved a comparable final loss after $50,000$ iterations. This indicates that while residual connections accelerate training, they yield equivalent accuracy at $N=6$.
Overall, the proposed method achieved faster convergence while maintaining a comparable final loss relative to its non-residual counterpart at shallow networks.

Fig.~\ref{totalloss}(b) shows the results for the deeper models ($N=18$). The total loss $\mathcal{L}$ successfully converged for all methods. Consistent with the $N=6$ case, PINNs-SIREN-Res achieved the fastest and most stable convergence, reaching the lowest loss among the compared models. Furthermore, within each activation function, the residual variants converged faster and attained lower final losses than their non-residual counterparts. Overall, the convergence trends observed at $N=6$ were preserved at $N=18$.

From these results, convergence was achieved for all methods, even in the deeper setting. Across both depths, the SIREN-based models consistently demonstrated faster convergence. Specifically, in the deeper setting, SIREN-based models achieved lower loss values than tanh-based models, and incorporating residual connections further improved performance, particularly in terms of convergence speed and the final loss.

\subsubsection{Estimation Accuracy over the Entire Evaluation Region}

To clarify the effect of network depth on estimation performance, we compare the mean NMSE across all evaluation points. Fig.~\ref{totalloss}(c) shows the relationship between NMSE and the number of hidden layers ($N=6, 10, 14, 18$), and Table~\ref{NMSE} lists the NMSE of the lowest-NMSE model across the tested depths for each method.

First, for the tanh-based models, PINNs-tanh showed little improvement as depth increased: the NMSE was approximately $-1.1$ dB at $N=6$ and $-2.6$ dB at $N=18$. 
In contrast, PINNs-tanh-Res improved markedly with depth, reaching an overall NMSE of approximately $-6.7$ dB at $N=18$, which is a 4.1 dB improvement over PINNs-tanh. 
Moreover, increasing the depth from $N=6$ to $N=18$ improved the NMSE of PINNs-tanh-Res by approximately $3.9$ dB.
These results indicate that residual connections are important for effectively exploiting increased depth in tanh-based PINNs.

Second, the results indicate that PINNs-SIREN achieves higher estimation accuracy than the tanh-based models among the compared models. For the best-performing models (across depths) of each type, PINNs-SIREN improved the NMSE by approximately $9.9$ dB compared with PINNs-tanh and by $5.8$ dB compared with PINNs-tanh-Res.
As shown in Fig.~\ref{totalloss}(c), the NMSE of PINNs-SIREN improved as the depth increased up to $N=14$; however,  further increasing the depth to $N=18$ led to a slight decrease in performance (approximately $0.2$ dB). This suggests that PINNs-SIREN outperforms tanh-based models; however, its estimation accuracy may degrade beyond a certain depth.

Next, we evaluate the proposed method, PINNs-SIREN-Res. As summarized in Table~\ref{NMSE}, the best-performing model (across depths) achieves NMSE improvements of approximately $11.6$ dB over PINNs-tanh, $7.5$ dB over PINNs-tanh-Res, and $1.7$ dB over PINNs-SIREN. Notably, while PINNs-SIREN showed a slight degradation when the depth was increased from $N=14$ to $N=18$, the proposed method instead improved the NMSE by 0.4 dB.
These results indicate that the best-performing proposed model outperforms the corresponding best-performing conventional models, and that its estimation accuracy improves with increasing depth up to 
$N=18$.

To assess the noise-reduction effect, Fig.~\ref{totalloss}(d) shows the mean NMSE at the measurement points. For all methods except PINNs-SIREN, the estimation accuracy improved with increasing depth.
In contrast, PINNs-SIREN shows a decrease in accuracy when the network is deepened to $N=18$. 
The proposed method improved the NMSE by approximately $14.9$ dB compared with PINNs-tanh, $10.5$ dB compared with PINNs-tanh-Res, and $3.0$ dB compared with PINNs-SIREN. Consistent with the trends observed at the evaluation points, these results indicate that the proposed method achieved the highest estimation accuracy at the measurement positions.

In the following sections, we compare the best-performing models for each method (in terms of network depth).

\begin{figure*}[tb]
\centering
\includegraphics[width=1\linewidth]{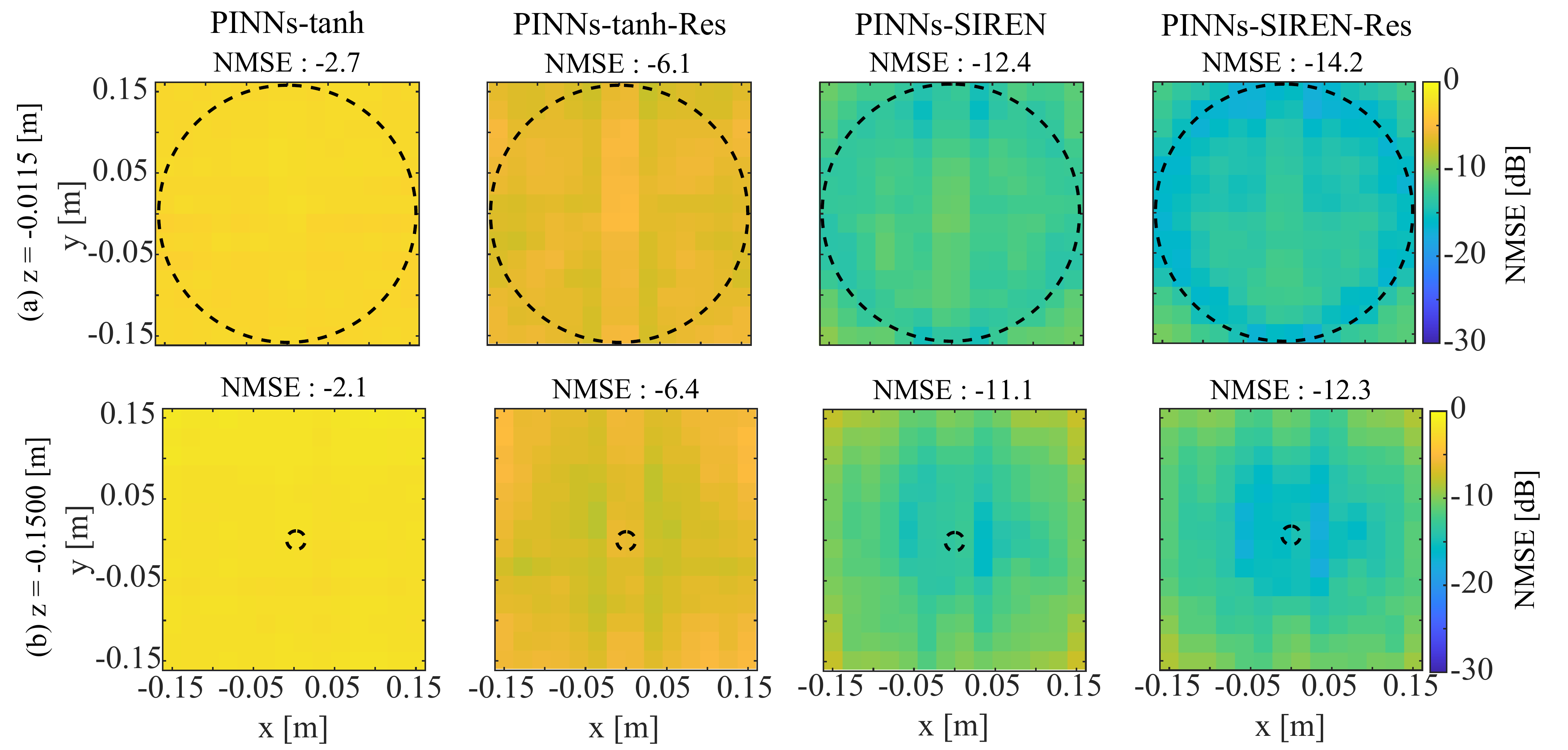}
\vspace{10pt}
\caption{Spatial distribution of NMSE[dB] on two x--y plane: (a) the center plane ($z = -0.0115$ m) and (b) the bottom plane ($z = -0.1500$ m) of the discretized estimation region. The black dashed lines indicate the boundaries of the measurement region.}
\vspace*{-3pt}
\label{NMSEMAP}
\end{figure*}
\begin{table*}[tb]
\centering
\caption{NMSE [dB] in the interpolation and extrapolation regions (inside and outside the measurement sphere, respectively) for each model.}
\setlength{\tabcolsep}{3pt}
\begin{tabular}{c|c|c|c|c}
\hline
Estimated region&
PINNs-tanh& 
PINNs-tanh-Res& 
PINNs-SIREN&
PINNs-SIREN-Res\\
\hline
Interpolation&
$-2.6$&
$-6.7$&
$-13.1$&
$-15.0$\\
Extrapolation&
$-2.5$&
$-6.7$&
$-12.1$&
$-13.7$\\
\hline
\end{tabular}
\label{NMSE_inout}
\end{table*}

\subsubsection{Frequency-Domain Accuracy}

Fig.~\ref{totalloss}(e) shows the NMSE as a function of frequency. The figure indicates that all methods achieve lower NMSE (i.e., higher accuracy) at low frequencies than at high frequencies.
In addition, fluctuations in NMSE are more pronounced in the low-frequency range across methods.

Compared with PINNs-tanh, PINNs-tanh-Res demonstrates improved estimation accuracy in the high-frequency range.  
The SIREN-based model further improves performance over the tanh-based models, and PINNs-SIREN-Res provides an additional gain, achieving the lowest NMSE overall.
These results suggest that the proposed method  improves estimation accuracy over a wide frequency range.

Table~\ref{NMSE_fq_block} summarizes the NMSE in four frequency bands: up to 1000~Hz, 1000--2000~Hz, 2000--3000~Hz, and 3000--4000~Hz.

PINNs-tanh attains an NMSE of -8.2~dB for frequencies up to 1000~Hz; however, its performance degrades substantially in higher bands compared with the other methods. For PINNs-tanh-Res, the NMSE increases by approximately $7.7$~dB from the sub-1000~Hz band to the 1000--2000~Hz band, but remains around $-5$~dB above 2000~Hz.

PINNs-SIREN achieves markedly better accuracy up to 1000~Hz (NMSE: $-21.2$~dB), but its NMSE gradually increases above 1000~Hz. The proposed PINNs-SIREN-Res achieves the best NMSE up to 1000~Hz ($-23.7$~dB) and shows less degradation than PINNs-SIREN at higher frequencies, indicating more stable estimation in the high-frequency bands.

Overall, the proposed method improves estimation accuracy across all frequency bands compared with the other methods. The general degradation at higher frequencies is likely due to shorter wavelengths, which make the wave field spatially more complex and therefore harder for the networks to resolve.

\begin{figure*}[tb]
\centering
\includegraphics[width=1\linewidth]{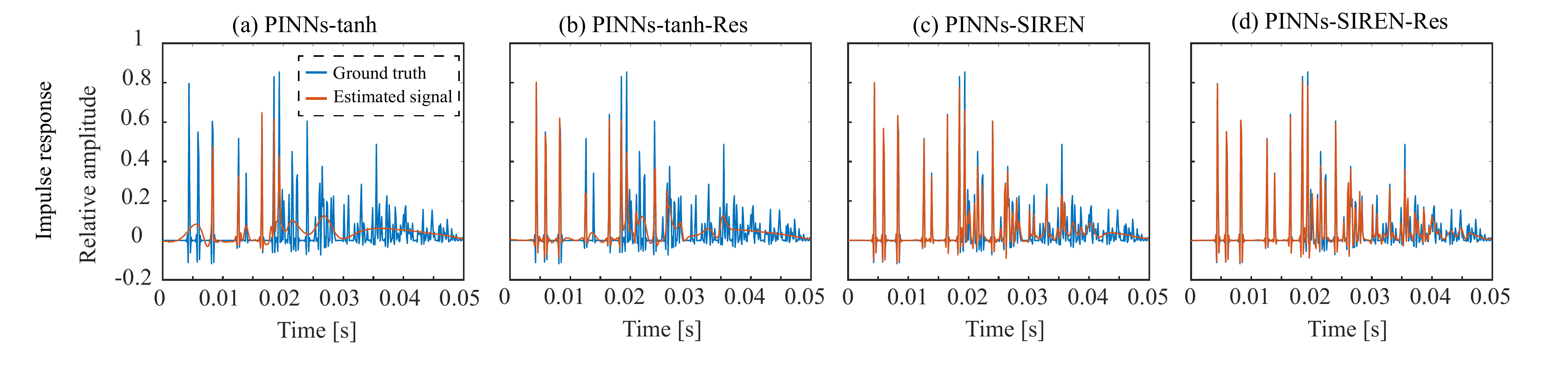}
\vspace{10pt}
\caption{Estimated impulse response and frequency response at $(x,y,z) =( -0.0115,-0.0115,-0.0115)$ m, comparing each model’s estimates with the ground truth.}
\vspace*{-3pt}
\label{press}
\end{figure*}

\begin{table*}[tb]
\centering
\caption{NMSE [dB] for the early and late segments of the RIR. This table lists the NMSE for each model for the early segment (direct sound and early reflections) and the late segment (late reverberation). The late segment is defined as the portion of the RIR after $0.025$~s.}
\setlength{\tabcolsep}{3pt}
\begin{tabular}{c|c|c|c|c}
\hline
Estimated region&
PINNs-tanh& 
PINNs-tanh-Res& 
PINNs-SIREN&
PINNs-SIREN-Res\\
\hline
Early segment&
$-2.5$&
$-8.3$&
$-17.1$&
$-17.5$\\
Late segment&
$-2.8$&
$-3.5$&
$-7.3$&
$-9.6$\\
\hline
\end{tabular}
\label{NMSE_ts}
\end{table*}

\subsection{Evaluation at Specific points}

In this section, we evaluate the estimated signals and NMSE at specific evaluation points to analyze (i) the spatial accuracy inside and outside the microphone array (i.e., in the interior and exterior regions relative to the array) and (ii) the temporal accuracy of the estimates.

\subsubsection{Interpolation and Extrapolation Performance}

We examined the spatial distribution of the estimation errors. Fig.~\ref{NMSEMAP} shows the spatial distribution of the NMSE over the evaluation points in the x--y plane.

Fig.~\ref{NMSEMAP}(a) corresponds to the plane at z$ = -0.0115$ m, which is the central slice of the evaluation domain along the z-axis. Note that, on this x--y plane, the measurement sphere is close to the boundary of the evaluation domain; consequently, many measurement points are concentrated near the periphery of the evaluation domain. Therefore, this slice is representative of interpolation performance. As shown in the figure, PINNs-tanh and PINNs-tanh-Res generally yielded a higher NMSE (i.e., lower accuracy) than the SIREN-based models across the evaluation points. The proposed method reduces the NMSE by approximately $11.5$~dB, $8.1$~dB, and $2.0$~dB compared with PINNs-tanh, PINNs-tanh-Res, and PINNs-SIREN, respectively.

Moreover, within each activation-function family, introducing residual connections improved the accuracy. Notably, PINNs-SIREN-Res achieved the lowest NMSE among all models, indicating a strong interpolation performance.

Fig.~\ref{NMSEMAP}(b) shows the evaluation points at z$= -0.1500$~m, corresponding to the bottommost layer of the evaluation domain. Because this layer lies on the boundary of the estimation domain, almost the entire plane falls within the extrapolation region; thus, this slice was used to evaluate the extrapolation accuracy.
Similar to Fig.~\ref{NMSEMAP}(a), PINNs-tanh and PINNs-tanh-Res yielded a higher NMSE (lower accuracy) than the SIREN-based models. Consistent with the results for the central plane (z$=-0.0115$~m), the proposed method achieved the best accuracy at this depth.


Table~\ref{NMSE_inout} summarizes the NMSE for the evaluation points located inside (interpolation region) and those outside (extrapolation region) the measurement points. In the interpolation region, the proposed method reduced the NMSE by approximately $12.4$~dB relative to PINNs-tanh, $8.3$~dB relative to PINNs-tanh-Res, and $1.9$~dB relative to PINNs-SIREN. In the extrapolation region, it reduced the NMSE by approximately $11.2$~dB, $7.0$~dB, and $1.6$ dB compared with these methods, respectively. These results suggest that incorporating residual connections improved the estimation accuracy in both interpolation and extrapolation.

\subsubsection{Time-Domain Analysis of Estimation Accuracy}

We evaluated the estimated RIRs in the time domain by focusing on direct sound and reflected components.
Fig.~\ref{press} shows a comparison of the RIRs and frequency responses at the $(x, y, z) = (-0.0115, -0.0115, -0.0115)$ m.
In Fig.~\ref{press}(a), PINNs-tanh failed to estimate the amplitude of the direct sound. As shown in Fig.~\ref{press}(b), although PINNs-tanh-Res captures the amplitude of the direct sound, it failed to capture the amplitude of the first-order reflections and was similar to PINNs-tanh and could not reliably estimate the subsequent reflections. 
In contrast, the SIREN-based models in Fig.~\ref{press}(c) and (d) capture the latter part of the RIR more effectively than tanh-based models. Moreover, a comparison of Figs.~\ref{press}(c) and (d) indicates that PINNs-SIREN-Res estimates the late-reverberation tail more accurately than PINNs-SIREN.

Despite adding white noise to the measured signals at an SNR of 20~dB, the RIR estimated by the proposed method closely matched the ground truth. Notably, the estimated signal did not exhibit white-noise-like fluctuations, indicating a low sensitivity to noise. This characteristic is particularly evident in the time range before $0.025$ s. These observations demonstrate the robustness of the proposed method against noise.

Table~\ref{NMSE_ts} presents the NMSE results segmented by time, distinguishing between the early and late portions of the signal (before and after $0.025$ s) at all the evaluation points. 
In the early segment, the proposed method improves the NMSE by approximately $15.0$~dB and $9.2$~dB relative to PINNs-tanh and PINNs-tanh-Res, respectively. Even compared with the strong baseline of PINNs-SIREN, an improvement of $0.4$~dB is achieved. 
In the late segment, the proposed method also maintains superior performance, improving the NMSE by approximately $6.8$ dB, $6.1$ dB, and $2.3$ dB compared with PINNs-tanh, PINNs-tanh-Res, and PINNs-SIREN, respectively.
Overall, these results indicate that SIREN-based models provide strong accuracy through the early-reflection region, and that residual connections further improve the estimation of late-reverberation components.

\section{Conclusion}
This study proposed a three-dimensional RIR estimation method based on PINNs with SIREN, incorporating deep hidden layers and residual connections (PINNs-SIREN-Res).
The experimental results demonstrate that the proposed method achieves higher estimation accuracy and improved robustness than models using other activation functions. In particular, the improvement in the accuracy of the reflection components was remarkable.

Furthermore, whereas PINNs-SIREN exhibited performance degradation when the number of layers was increased from 14 to 18, the proposed model continued to improve with increased depth. The proposed method also outperformed PINNs-SIREN in estimating reflection components. In the frequency domain, consistent improvements were observed across all frequency bands, although the estimation accuracy degraded as the frequency increased for all methods.

Future work includes refining the model to improve estimation accuracy over a broader spatial domain, enhancing performance at higher frequencies, and improving the estimation of late reverberation. In addition, although this study validated the proposed method through simulation experiments, extending it to real-world measurements and practical acoustic environments is an important direction for future research.

\end{document}